\documentclass[]{spie}  

\newcommand\arcsec{\mbox{$^{\prime\prime}$}}%
\newcommand\fdg{\mbox{$.\!\!^\circ$}}%
\newcommand\degr{\mbox{$^\circ$}}%

\newcommand\farcs{\mbox{$.\!\!^{\prime\prime}$}}%
\let\farcs\farcs 

\usepackage{amsmath,amsfonts,amssymb}
\usepackage{graphicx}
\usepackage[colorlinks=true, allcolors=blue]{hyperref}
\usepackage[nolist]{acronym}

\newacro{adi}[ADI]{angular differential imaging}
\newacro{ao}[AO]{adaptive optics}
\newacro{aolp}[AOLP]{angle of linear polarization}
\newacro{charis}[CHARIS]{Coronagraphic High Angular Resolution Imaging Spectrograph}
\newacro{de}[DE]{differential evolution}
\newacro{dpp}[DPP]{Data Processing Pipeline}
\newacro{drp}[DRP]{Data Reduction Pipeline}
\newacro{fov}[FOV]{field of view}
\newacro{fwhm}[FWHM]{full width at half maximum}
\newacro{gpi}[GPI]{Gemini Planet Imager}
\newacro{hiciao}[HiCIAO]{High-Contrast Coronographic Imager for Adaptive Optics}
\newacro{hwp}[HWP]{half-wave plate}
\newacro{ifs}[IFS]{integral field spectrograph}
\newacro{irdis}[IRDIS]{InfraRed Dual-band Imager and Spectrograph}
\newacro{nir}[NIR]{near-infrared}
\newacro{pdi}[PDI]{polarimetric differential imaging}
\newacro{pi}[PI]{polarized intensity}
\newacro{piaacmc}[PIAACMC]{phase-induced amplitude apodization complex mask coronagraph}
\newacro{psf}[PSF]{point spread function}
\newacro{rdi}[RDI]{reference star differential imaging}
\newacro{sb}[SB]{surface brightness}
\newacro{scexao}[SCExAO]{Subaru  Coronagraphic  Extreme  Adaptive  Optics}
\newacro{sdi}[SDI]{spectral differential imaging}
\newacro{snr}[SNR]{signal-to-noise ratio}
\newacro{sphere}[SPHERE]{Spectro-Polarimetric High-contrast Exoplanet REsearch instrument}
\newacro{vampires}[VAMPIRES]{Visible Aperture Masking Polarimetric Imager for Resolved Exoplanetary Structures}
\newacro{vapp}[vAPP]{vector Apodizing Phase Plate}
\newacro{vlt}[VLT]{Very Large Telescope}

\title{High-contrast integral field spectropolarimetry of planet-forming disks with SCExAO/CHARIS}

\author[a]{Kellen Lawson}
\author[b,c,d]{Thayne Currie}
\author[a]{John P. Wisniewski}
\author[e,b,f]{Jun Hashimoto}
\author[b,g,e]{Olivier Guyon}
\author[h]{N. Jeremy Kasdin}
\author[i]{Tyler D. Groff}
\author[b]{Julien Lozi}
\author[j]{Timothy D. Brandt}
\author[k]{Jeffrey Chilcote}
\author[b]{Vincent Deo}
\author[l,m]{Taichi Uyama} 
\author[b]{Sebastien Vievard}

\affil[a]{Department of Physics and Astronomy, University of Oklahoma, Norman, OK, USA}

\affil[b]{Subaru Telescope, National Astronomical Observatory of Japan, 650 North A`oh$\bar{o}$k$\bar{u}$ Place, Hilo, HI, USA}

\affil[c]{NASA-Ames Research Center, Moffett Blvd., Moffett Field, CA, USA}

\affil[d]{Eureka Scientific, 2452 Delmer Street Suite 100, Oakland, CA, USA}

\affil[e]{Astrobiology Center, National Institutes of Natural Sciences, 2-21-1, Osawa, Mitaka, Tokyo, 181-8588, Japan}

\affil[f]{Department of Astronomy, School of Science, Graduate University for Advanced Studies (SOKENDAI), Mitaka, Tokyo 181-8588, Japan}

\affil[g]{College of Optical Sciences, University of Arizona, Tucson, AZ, USA}

\affil[h]{Department of Mechanical Engineering, Princeton University, Princeton, NJ, USA}

\affil[i]{NASA-Goddard Space Flight Center, Greenbelt, MD, USA}

\affil[j]{Department of Physics, University of California, Santa Barbara, Santa Barbara, California, USA}

\affil[k]{Department of Physics, University of Notre Dame, South Bend, IN, USA}

\affil[l]{Infrared Processing and Analysis Center, California Institute of Technology, 1200 E. California Boulevard, Pasadena, CA 91125, USA}

\affil[m]{NASA Exoplanet Science Institute, Pasadena, CA 91125, USA}

\authorinfo{Further author information: (Send correspondence to K. Lawson)\\K. Lawson: E-mail: kellenlawson@gmail.com}

\begin{document} 
\maketitle

\begin{abstract}
We describe a new high-contrast imaging capability well suited for studying planet-forming disks: \ac{nir} high-contrast spectropolarimetric imaging with the \ac{scexao} system coupled with the \ac{charis} \ac{ifs}. The advent of extreme \ac{ao} systems, like SCExAO, has enabled recovery of planet-mass companions at the expected locations of gas-giant formation in young disks alongside disk structures (such as gaps or spirals) that may indicate protoplanet formation. In combination with \ac{scexao}, the \ac{charis} \ac{ifs} in polarimetry mode allows characterization of these systems at wavelengths spanning the \ac{nir} J, H, and K bands ($1.1-2.4$ $\mu m$, $R\sim20$) and at angular separations as small as 0$\farcs$04. By comparing the resulting images with forward-modeled scattered light or 3D radiative-transfer models, the likely origins of any observed features can be assessed. Utilization of swift optimization algorithms, such as \ac{de}, to identify model parameters that best reproduce the observations allows plausible disk geometries to be explored efficiently. The recent addition of \ac{charis}’s unique integral field spectropolarimetry mode has further facilitated the study of planet-forming disks — aiding in the confirmation of candidate protoplanets, the diagnosis of disk structures, and the characterization of dust grain populations. We summarize preliminary results for two young planet-forming disk systems based on observations with the novel integral field spectropolarimetry mode for \ac{scexao}/\ac{charis}.\acresetall
\end{abstract}


\section{INTRODUCTION}\label{sec:intro} 

Protoplanetary disks serve as benchmark systems to study how and where exoplanets form. With the advent of ground-based extreme \ac{ao} facilities, such as \ac{sphere} \cite{Beuzit2019}, \ac{gpi} \cite{Macintosh2015}, and \ac{scexao} \cite{Jovanovic2015,Lozi2018,Currie2020spie}, we have entered a revolutionary era in the study of young protoplanetary disks. It is now feasible to both a) spatially resolve the morphological signatures within disks that could be caused by newly formed/forming sub-stellar objects and planets (such as gaps or spirals), and b) identify young candidate planets and sub-stellar objects that may cause them (e.g. Refs \citenum{Lagrange2010, Keppler2018}).

Differential imaging techniques for high-contrast imaging in total intensity, such as \ac{adi}\cite{Marois2006}, \ac{sdi}\cite{SparksFord2002}, and \ac{rdi}, attempt to disentangle the light of the bright stellar \ac{psf} from that of any circumstellar sources by modeling and subtracting the starlight in a data sequence. While these techniques are generally reasonably well suited for the recovery and characterization of isolated point-like companion candidates, the presence of signal from circumstellar disks can significantly inhibit their efficacy -- resulting in products in which disk signal is substantially attenuated. For \ac{adi} and \ac{sdi}, this results from self-subtraction (where disk signal is erroneously included in the model \ac{psf}) and oversubtraction (where the presence of disk signal in the data results in a brighter \ac{psf} model than if it were not present)\cite{Pueyo2016,Currie2019b}. While \ac{rdi} is expected to perform better for disk imaging -- by eliminating the possibility of self-subtraction -- oversubtraction still occurs, and can be significant for bright disks or disks which dominate the \ac{fov}. As a result, any analysis of disks recovered from total intensity differential imaging will generally require the attenuation to be quantified. Further, rigorous assessment of planet candidates embedded in disk material requires a means by which to differentiate a planet's signal from that of disk substructure (e.g. Refs \citenum{Rich2019, Currie2019}). Forward-modeling of synthetic disk models is a common approach for approximating disk attenuation (e.g. Refs \citenum{Currie2019, Lawson2020}). Combined with integral field spectroscopy or multi-wavelength imaging, this can also provide a means by which to validate or invalidate candidate embedded planets. However, given the complexity of total intensity \ac{psf}-subtraction and of the geometry of many protoplanetary disks, forward-modeling can be prohibitively time-consuming.

High-contrast \ac{nir} polarimetric imagers -- such as Gemini South's \ac{gpi}\cite{Macintosh2015, Perrin2015}, VLT's \ac{sphere}-\ac{irdis}\cite{deBoer2020, vanHolstein2020_irdis}, and Subaru's \ac{hiciao}\cite{Suzuki2010} -- have been prolific in detecting and characterizing circumstellar disks through the use of another differential imaging technique: \ac{pdi}. While \ac{pdi} is limited to the recovery of polarized flux, it provides an important complement for total intensity differential imaging by producing a nearly unattenuated view of the disk. Additionally, since self-luminous sub-stellar objects and exoplanets are expected to be unpolarized while light from spatially resolved circumstellar disks is $25-50\%$ polarized \cite{Perrin2009, Uyama2017}, comparison of total and polarized intensity imaging can serve as another powerful tool for disentangling planet signals from disk sub-structure. Since the \ac{pdi} procedure is computationally simple, forward modeling disks to simulate the effects of \ac{pdi} is substantially less time consuming -- thus allowing for more detailed investigation of disk model parameters. 

Since the decommissioning of Subaru's \ac{hiciao}, the \ac{charis} \ac{nir} \ac{ifs} \cite{Groff2016} (paired with \ac{scexao}) has served as Subaru's \ac{nir} total intensity high-contrast imager. More recently, \ac{charis} was upgraded to include spectropolarimetric capabilities as well\cite{Lozi2020}. Unprecedented among extreme \ac{ao} imagers, this new spectropolarimetric imaging mode allows polarimetric observations to be conducted at the same array of wavelengths as its classical (total intensity) observing mode. This capability makes accessible many novel investigations relevant for protoplanetary disk studies, such as characterization of the wavelength dependence of the scattering phase function in disks.

We present preliminary science products from the novel integral field spectropolarimetry mode, or \ac{pdi}-mode, for Subaru's \ac{scexao}/\ac{charis}. Combining the benefits of multi-wavelength \ac{ifs} imaging with those of polarimetric imaging, this new observing mode provides an exciting tool for groups studying circumstellar disks. Herein, we: 1) provide a summary of the \ac{charis} \ac{pdi}-mode, outline its available configurations, and make recommendations for those observing with this mode (Section \ref{sec:configs}), 2) provide a description of the data processing and calibration procedures for \ac{charis} \ac{pdi} data, implemented in a new module for the \ac{charis} \ac{dpp} (Section \ref{sec:data_processing}), 3) present preliminary science products for two protoplanetary disk targets observed with \ac{charis} in \ac{pdi}-mode (Section \ref{sec:results}), 4) demonstrate disk forward-modeling for \ac{charis} \ac{pdi} data (Section \ref{sec:fwdmodeling}), and 5) conclude by summarizing compelling future use-cases for \ac{charis}'s \ac{pdi}-mode in application to the study of circumstellar disks (Section \ref{sec:conclusion}).

\section{CHARIS Spectropolarimetry Mode}\label{sec:configs}
    
    The spectropolarimetric observing mode for \ac{charis} is enabled by the addition of a field stop and Wollaston prism upstream of the imager in combination with the existing \ac{hwp} -- originally added to enable polarimetry with \ac{hiciao}. The  Wollaston prism serves to split incoming light to two orthogonally polarized states, while the \ac{hwp} enables shifting of the polarization direction being measured, and the field stop eliminates cross-talk between the two polarizations\footnote{see Refs \citenum{vanHolstein2020, Lozi2020} for a more complete description of the instrumental design of the \ac{charis} spectropolarimetry mode}.  Together, the Wollaston prism and \ac{hwp} enable measurement of linear polarization parameters (Stokes-$Q$ and $U$; in addition to the total intensity, Stokes-$I$, measured by the classical-mode), at \ac{charis}'s typical array of wavelengths, at the cost of reducing \ac{charis}'s \ac{fov} from $\sim 2\arcsec{} \times 2\arcsec{}$ to $\sim 1\arcsec{} \times 2\arcsec{}$. In the typical \ac{pdi} procedure, circumstellar polarized flux is isolated by making observations at \ac{hwp} angles of 0\degr{} and 45\degr{} or 22\fdg{}5 and 67\fdg{}5 and using ``double-differencing'' (see a full description of \ac{pdi} in Section \ref{sec:pdi}).
    
    A complete model for the correction of instrumental polarization for \ac{charis} \ac{pdi}-mode is near completion (see Ref \citenum{vanHolstein2020}) and will be implemented in the data processing pipeline for \ac{charis} soon after. In lieu of this correction, results herein should be considered preliminary.
    
    \subsection{Half-wave Plate Cycles}
    
    Switching of the \ac{hwp} requires $\sim 6$ seconds. While the most straight-forward strategy might be to change the \ac{hwp} angle between each exposure, this could introduce significant overheads for targets requiring short exposure times. Especially in these cases, taking multiple exposures at a given \ac{hwp} position before each switch may improve the quality of results by enabling more time on target overall. This introduces two possible costs, however. The first is that the suppression of unpolarized stellar flux may be compromised by evolution of the diffraction pattern over the relatively protracted \ac{hwp} cycles. Additionally, since \ac{scexao} operates exclusively in pupil-tracking mode, exposures at paired \ac{hwp} positions for double differencing will be at slightly different parallactic angles. This could result in noticeable smearing of circumstellar features for targets observed near zenith (though the impact of this is unlikely to be significant, see: Section \ref{sec:fwdmodeling}). As such, target properties, observing conditions, and the circumstellar features of interest should all be considered when planning observations.
    
    \subsection{Instrument Configurations}
    
    With some exceptions, the available configurations for \ac{charis}'s \ac{pdi}-mode are the same as for its classical-mode. These configurations are summarized hereafter.
    
    \textbf{Astrogrid} -- As with \ac{charis}'s classical imaging mode, \ac{charis} \ac{pdi}-mode allows for use of the astrogrid during observations. The astrogrid introduces a set of satellite spots which are attenuated copies of the unocculted stellar \ac{psf}. The astrogrid enables high quality image registration and flux calibration even when the star is obscured by a coronagraph. While the same array of astrogrid spacings are available as for \ac{charis}'s classical mode, the standard classical-mode spacing of 15.9 $\lambda / D$ places spots outside of the narrower $2\arcsec{} \times 1 \arcsec{}$ \ac{fov} of \ac{pdi}-mode by the latter channels of H-band. Thus a spacing of $\sim$ 11.2 $\lambda / D$ or smaller should be utilized for the \ac{pdi}-mode \cite{Lozi2020}. While a limited sequence of exposures utilizing the astrogrid might be sufficient for reasonable registration and flux calibration, use of the astrogrid throughout the entirety of an observing sequence is strongly recommended unless a compelling science case exists to preclude this -- such as simultaneous use of the \ac{vampires} instrument, which cannot be used with the astrogrid on. 
    
    \textbf{Filters and Spectral Resolution} -- $J$-band ($1.18-1.33$ $\mu m$), $H$-band ($1.49-1.78$ $\mu m$), and $K$-band ($2.02 - 2.38$ $\mu m$) are available with high spectral resolution ($R\sim 70-90$, yielding $15-20$ wavelength channels in extracted image cubes).  \ac{charis} broadband ($1.15-2.39$ $\mu m$), and Cold ND3 ($1.17 - 2.37$ $\mu m$) are available with low spectral resolution ($R\sim19$, yielding 22 wavelength channels in extracted image cubes).
    
    \textbf{Coronagraphs} -- the Lyot coronagraphs\footnote{See \url{https://www.naoj.org/Projects/SCEXAO/} for information regarding available coronagraph sizes.} can be used with \ac{charis} \ac{pdi}-mode for any available filter. Additionally, the \ac{piaacmc} can be used when observing in $H$-band. Other coronagraphs that are generally available for \ac{scexao}, e.g. \ac{vapp} and vortex coronagraphs, induce intractable changes in the polarization state of incident light and are thus unsuitable for observations in \ac{pdi}-mode.
    
    \subsection{Calibration Data}
    With the complete Mueller matrix model for instrumental polarization implemented, observations of polarized/unpolarized standard stars will not be directly necessary for calibration of science products. However, these observations are still recommended, as they require very little time, can be used to verify the integrity of the instrumental polarization corrections, and can be useful in \ac{psf}-subtraction of total-intensity products by enabling \ac{rdi}.
    
    \subsection{Observations}\label{sec:obs}
    To demonstrate the use of the methods described herein and to showcase this new observing mode, we include preliminary results of observations of two protoplanetary disk-bearing stars. In both cases, we reserve detailed analysis, discussion, and target-specific conclusions for future publications.
    
    \textbf{AB Aurigae}-- 
    AB Aurigae (1--3 Myr \cite{Kenyon2008}, $d=156$ pc\cite{Gaia2016, Gaia2021}) is a pre-main sequence star that hosts a complex protoplanetary disk bearing spiral arms at scales from tens of au\cite{Hashimoto2011, Boccaletti2020} to hundreds\cite{Grady1999, Fukugawa2004}.

    On 2020 October 4, we observed AB Aur with \ac{scexao}/\ac{charis} in low-res ($R \sim 19$), broadband ($1.15 - 2.39$ $\mu m$), \ac{pdi}-mode, using a 113 mas radius Lyot coronagraph and a four-spot astrogrid of spacing $11.2$ $\lambda / D$. \ac{ao} performance was average / below-average, with a strong wind-driven halo. We collected 73 science exposures of 60.48 seconds each for a combined integration time of $\sim 73.6$ minutes, achieving a total field rotation of $79\fdg{}95$. Sky frames were also collected at the end of the sequence.
    
    Additional observations were made earlier in the same night using an astrogrid spacing of $15.9$ $\lambda / D$. Since the satellite spots are not recoverable in every channel at this spacing, image calibration for this data is much less precise. As such, we present only the observations with $11.2$ $\lambda / D$ astrogrid spacing herein.
    
    \textbf{TW Hydrae}-- TW Hydrae (3--10 Myr \cite{Hoff1998, Barrado2006, Vacca2011, vanBoekel2017}, $d=60$ pc\cite{Gaia2016, Gaia2021}) is a nearby T-Tauri star with a nearly face-on disk which shows multiple radial gaps that could suggest the presence of planets or their ongoing formation \cite{Akiyama2015, vanBoekel2017}.
    
    On 2021 March 20, we observed TW Hya with \ac{charis} in the same configuration as for AB Aur, but using a smaller Lyot coronagraph (78 mas radius). The \ac{ao} performance for TW Hya was worse than that of AB Aur, largely as a result of TW Hya's fainter apparent magnitude and lower altitude ($\sim35\degr{}$ versus $\gtrsim 74\degr$ for AB Aur). We collected 52 science exposures of 60.48 seconds each for a combined integration time of $\sim 52.4$ minutes, achieving a total field rotation of $15\fdg{}84$. No sky frames were collected with these data.
    
\section{CHARIS PDI Data Reduction and Processing}\label{sec:data_processing}
    \subsection{Cube Extraction}
\begin{figure}
    \includegraphics[width=\textwidth]{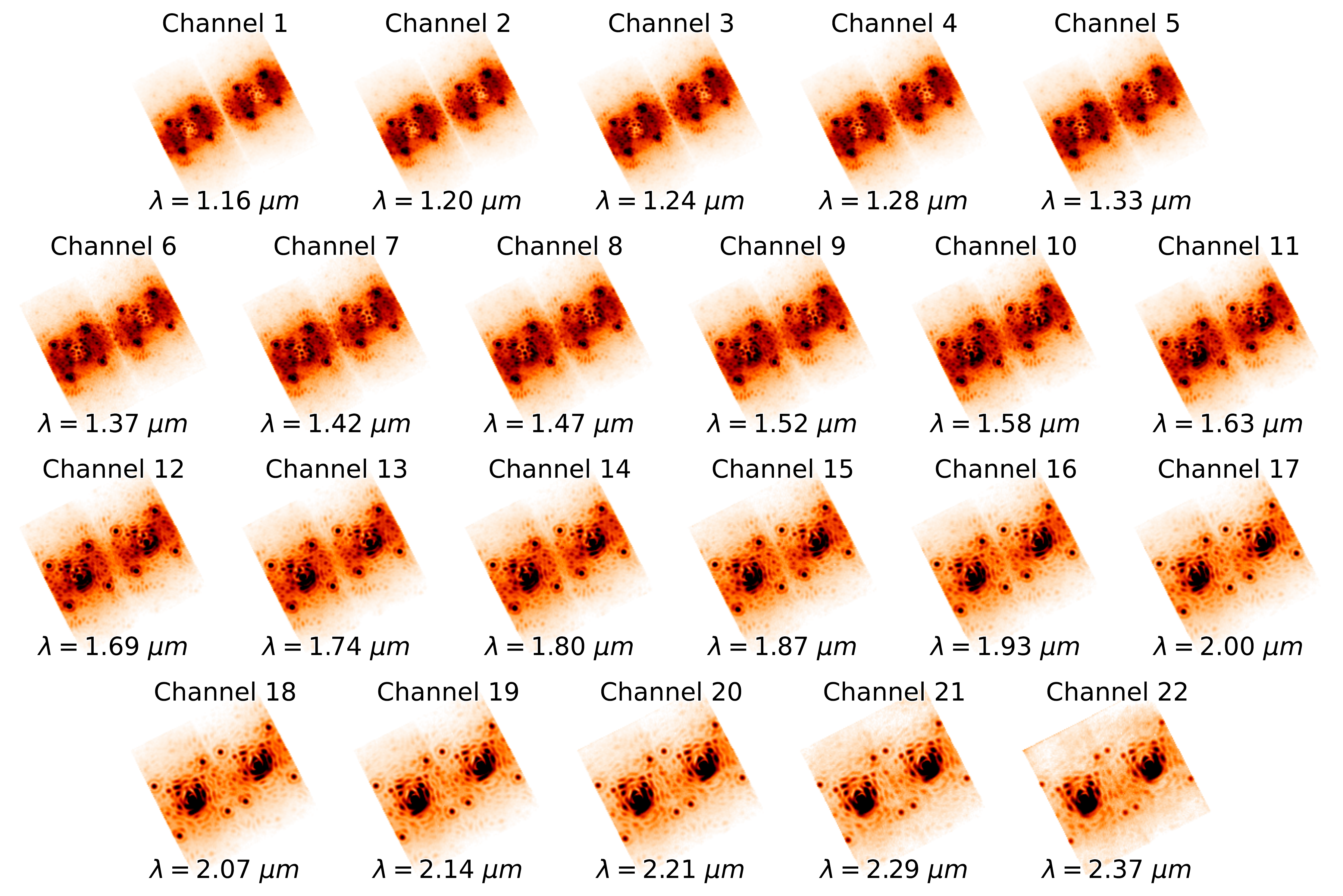}
    \caption{The 22 individual wavelength channels of an extracted  low-res, broadband, \ac{pdi}-mode \ac{charis} exposure of AB Aur -- showing both left and right polarizations. For this exposure, a four-satellite-spot astrogrid with spacing of 11.2 $\lambda / D$, and a Lyot-coronagraph of radius 113 mas were utilized.
    \label{fig:raw_channels}
    }
    \end{figure}

    \ac{charis} \ac{pdi} data were extracted from raw \ac{ifs} exposures using the \ac{charis} \ac{drp} \cite{Brandt2017} with modifications to read noise suppression provided in Ref \citenum{Currie2020spie}. The 22 wavelength channels of a single extracted AB Aur image cube are shown in Figure \ref{fig:raw_channels}.
    
    \subsection{Preprocessing}
    Preliminary processing of data for \ac{pdi}-mode -- e.g. sky subtraction, image registration, and spectrophotometric calibration --  is carried out largely as typical for \ac{charis} classical-mode data \cite{Currie2018}, with the following caveats.
    
    \begin{enumerate}
        \item \textbf{Registration} -- the left and right polarized halves of each image cube are first split into separate files. The sets of left and right frames are then registered independently, but otherwise as standard for \ac{charis} data (e.g. using satellite spots for occulted data, or \ac{psf} centroiding for unocculted data).
        \item \textbf{Spectrophotometric Calibration} -- for this purpose, the sets of registered left and right polarization cubes are added to form a sequence of single sum total intensity images. These images are then flux calibrated in the same manner as classical-mode data -- resulting in a set of scaling factors for each of the 22 wavelength channels. These scaling factors are then applied to the original constituent left and right polarized image cubes.
    \end{enumerate}

    \subsection{Half-Wave Plate Cycle Matching}

    To enable \ac{pdi} using double-differencing, we group observations into sets containing exposures at each of the four \ac{hwp} positions. While sets of two images (with \ac{hwp} angles of 0\degr{} and 45\degr{} or 22\fdg{}5 and 67\fdg{}5) would be sufficient to enable double-differencing (see Section \ref{sec:pdi}), forming groups of complete \ac{hwp} cycles allows polarized intensity outputs for each individual cycle -- which can be helpful in evaluating the validity of marginally detected features in products for the fully-combined observing sequence.
    
    Here, each exposure is used only once, and is grouped to minimize parallactic angle differences between the exposures in a cycle. Ultimately, this procedure results in a number of full \ac{hwp} cycles equal to the minimum number of observations for any one \ac{hwp} position after low-quality or otherwise problematic exposures are eliminated. Any unmatched exposures are not utilized in the \ac{pdi} procedure.

    \subsection{Polarimetric Differential Imaging}\label{sec:pdi}
\begin{figure}
    \includegraphics[width=\textwidth]{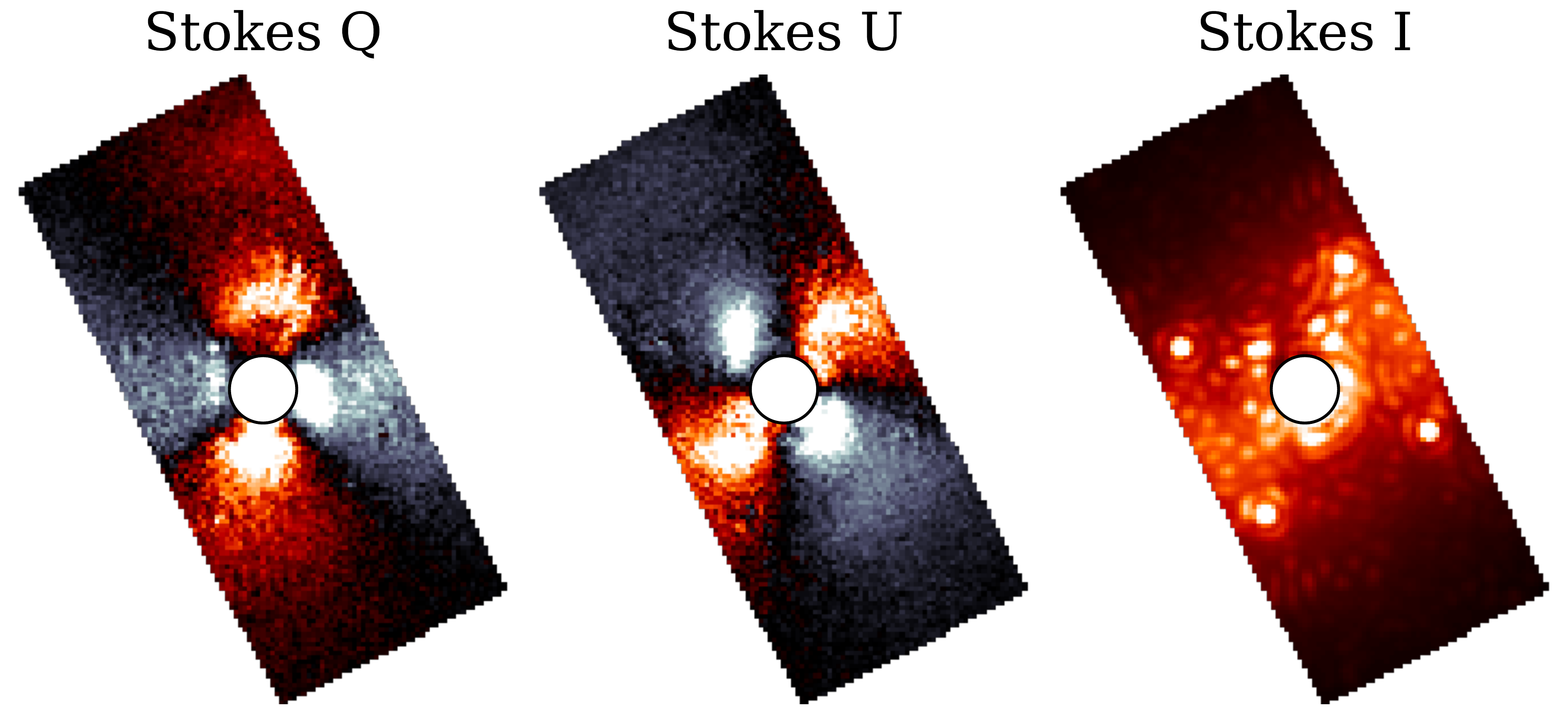}
        \caption{A single wavelength channel (channel 11, $\lambda = 1.63$ $\mu m$) showing Stokes $Q$, $U$, and $I$ images resulting for one \ac{hwp} cycle of \ac{charis} AB Aur data (including the exposure of Figure \ref{fig:raw_channels}). $I$ is computed from the $I_Q$ and $I_U$ products for the cycle as $I=0.5(I_Q+I_U)$. $Q$ and $U$ images have a first order instrumental polarization correction applied. Color stretches are symmetric about zero, with the range for $I$ differing from that of $Q$ and $U$ for visualization. The frames have not yet been derotated.
    \label{fig:stokes_params}
    }
    \end{figure}

    Reduction of the preprocessed data sequence to final polarimetry products is carried out using the typical double differencing procedure for each \ac{hwp} cycle (see e.g. Ref \citenum{deBoer2020}). First, single sums and differences are computed for each exposure (adopting the notation of Ref \citenum{vanHolstein2020}):
    \begin{equation}\label{eq:single_diff}
        X^\pm = I_{det,L} - I_{det,R} \, ,
    \end{equation}
    \begin{equation}\label{eq:single_sum}
        I_{X^\pm} = I_{det,L} + I_{det,R} \, ,
    \end{equation}
    where $I_{det,L}$ and $I_{det,R}$ are the left and right images respectively, $X^\pm$ are the single differences -- $Q^+$, $U^+$, $Q^-$, $U^-$ -- and $I_{X^\pm}$ are the single sums -- $I_{Q^+}$, $I_{U^+}$, $I_{Q^-}$, $I_{U^-}$ -- for exposures at \ac{hwp} angles of 0\degr{}, 22\fdg{}5, 45\degr{}, and 67\fdg{}5 respectively. From these, double sums and differences are computed as:
    
    \begin{equation}\label{eq:double_diff}
        X = \frac{1}{2}(X^+ - X^-) \, ,
    \end{equation}
    
    \begin{equation}\label{eq:double_sum}
        I_X = \frac{1}{2}(I_{X^+} + I_{X^-}) \, ,
    \end{equation}
    
    where the double difference, $X$, corresponds to Stokes $Q$ or $U$, and the double sum, $I_X$, corresponds to the components of Stokes $I$: $I_Q$ or $I_U$. At this stage, in lieu of a final model for instrumental polarization, we make a first order approximation of instrumental polarization following the procedure of Ref \citenum{deBoer2020} and summarized hereafter. For each double difference image cube,  $X$, resulting from the procedure above, the ratio with the corresponding double-sum image cube, $I_X$, is computed: $x = X / I_X$. Then, in each wavelength channel, $i$, we calculate the median value of $x$ within a circular stellocentric aperture (excluding the inner occulted region if a coronagraph was utilized): $c_X^i$.  The first-order instrumental polarization subtracted cube is then:
    \begin{equation}\label{eq:double_diff_ips}
        X_{\rm{IPS}} = X - c_X \cdot I_X \, ,
    \end{equation}
    where $c_X$ is the array of coefficients for the $N_\lambda$ wavelength channels, $c_X = [c_X^1, c_X^2, ..., c_X^{N_\lambda}]$. Since this procedure assumes that the polarization measured in the aperture should be zero in the absence of instrumental polarization, the coverage of the aperture should be chosen to minimize inclusion of circumstellar material. An example of a single wavelength channel of Stokes parameter images resulting from this procedure is shown in Figure \ref{fig:stokes_params}.
    
    Once $Q_{\rm{IPS}}$ and $U_{\rm{IPS}}$ (hereafter simply $Q$ and $U$) are attained, the products can be derotated to a ``north-up" orientation. Noting that a) $X^+$ and $X^-$ are not perfectly contemporaneous, and b) \ac{scexao} observes exclusively in pupil-tracking mode: the constituent images will have slightly different angular offsets from north-up. Here, we simply use the average of the parallactic angles of the two constituent frames to compute the rotation needed for a north-up orientation. The effect of this difference in parallactic angle is assessed in Section \ref{sec:fwdmodeling}.
    
    From the north-up polarization products, we can compute a few distinct polarized intensity products. Generally:
    \begin{equation}\label{eq:pi}
        \rm{PI} = \sqrt{Q^2+U^2} \, ,
    \end{equation}
    where $\rm{PI}$ is the polarized intensity. For a single \ac{hwp} cycle:
    \begin{enumerate}
        \item a corresponding $\rm{PI}$ cube can be created,
        \item a selection of wavelength slices of the $Q$ and $U$ cubes can be combined to approximate common \ac{nir} filters (e.g., $J$, $H$ or $K$ bands), from which a corresponding $\rm{PI}$ image can be calculated,
        \item all wavelength slices for $Q$ and $U$ cubes can be combined to enable calculation of a higher \ac{snr} $\rm{PI}$ image.
    \end{enumerate} 
    
    Additionally, we can average the full set of $Q$ and $U$ image cubes over all cycles to produce a sequence-combined $Q$ and $U$ image cube. From this, the same array of options for presenting $\rm{PI}$ are available. Since the nominal $1\arcsec{} \times 2 \arcsec{}$ \ac{fov} has been derotated by varying amounts, a sequence-combined product will have full coverage over $r \lesssim 0\farcs{}5$ and partial coverage of a larger area -- approaching a $\sim 1 \arcsec{}$ circle of partial coverage when a total field rotation of $\Delta \rm{PA} \sim 125\degr{}$ is achieved. For $\Delta \rm{PA} \lesssim 125\degr{}$, the region of zero coverage in a sequence-combined product manifests as two opposing wedges extending from $r\gtrsim 0\farcs{}5$ (see Figure \ref{fig:coverage_maps}). We note, however, that the exact \ac{fov} recovered will depend on the details of the dataset.

    For each of these cases, we can also calculate the \ac{aolp}:
    \begin{equation}\label{eq:aolp}
        \rm{AOLP} = \frac{1}{2} \arctan{\left(\frac{U}{Q}\right)} \, ,
    \end{equation}
    
    as well as the azimuthal Stokes parameters:
    \begin{equation}\label{eq:qphi}
        Q_\phi = -Q \cos{\left(2\phi\right)} - U \sin{\left(2\phi\right)}\, ,
    \end{equation}
    \begin{equation}\label{eq:uphi}
        U_\phi = Q \sin{\left(2\phi\right)} - U \cos{\left(2\phi\right)}\, ,
    \end{equation}
    where $\phi$ is the pixel-wise azimuthal angle relative to the image center.

\section{Preliminary CHARIS Spectropolarimetry Results}\label{sec:results}
\begin{figure}
        \includegraphics[width=\textwidth]{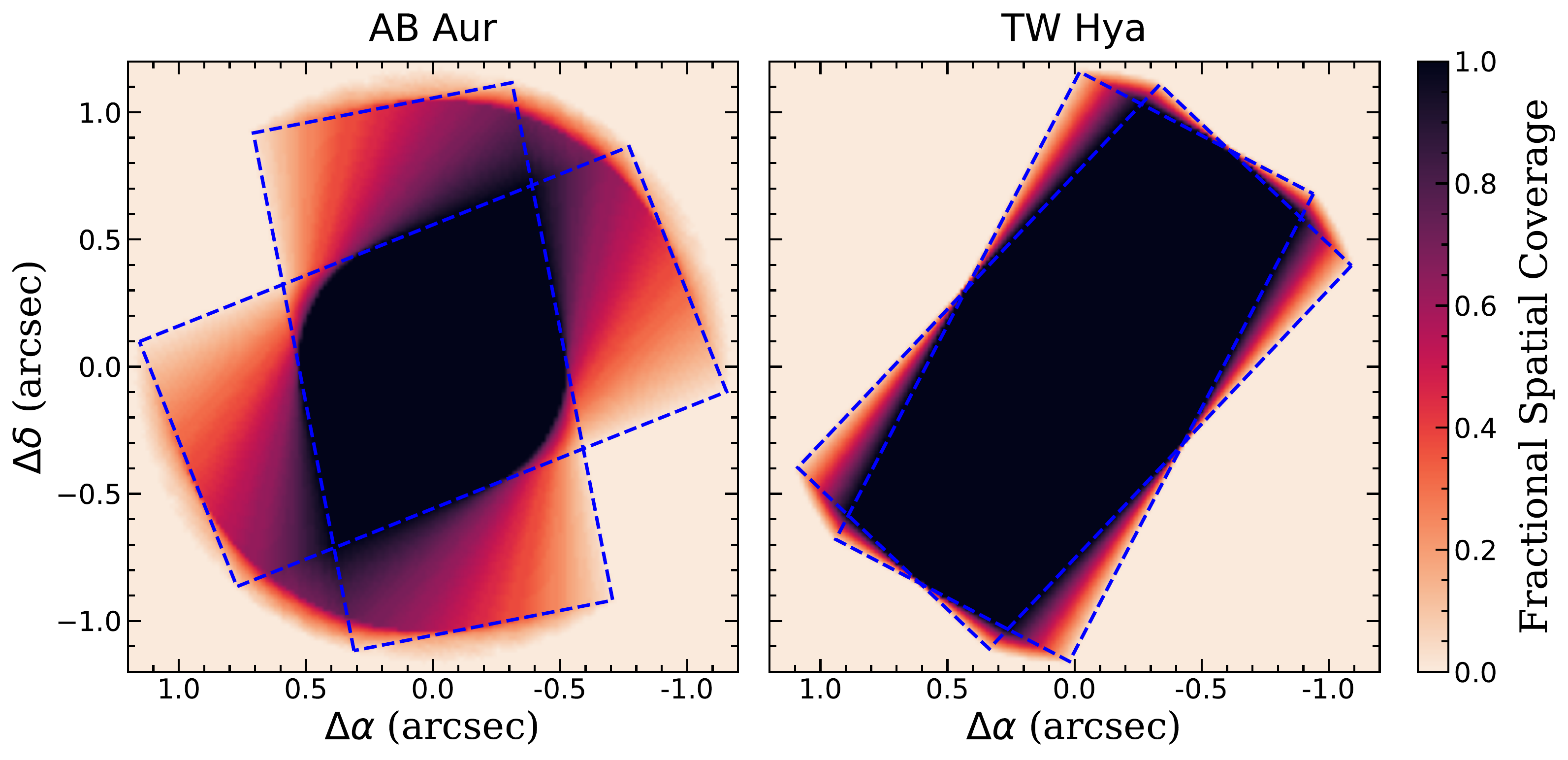}
            \caption{Approximate spatial coverage maps for our sequence-combined AB Aur and TW Hya products. The dashed blue rectangles indicate the extents of the derotated first and last image in each sequence. TW Hya has significantly more area with zero coverage but also more area with full coverage as a result of the TW Hya data's lesser field rotation.
        \label{fig:coverage_maps}
        }
    \end{figure}

    We apply the methods of Section \ref{sec:data_processing} to the \ac{charis} \ac{pdi} data of AB Aur and TW Hya (see Section \ref{sec:obs}). Both targets were registered and flux calibrated using the induced astrogrid. Approximate coverage maps for the final sequence-combined products are visualized in Figure \ref{fig:coverage_maps}. 
    
   For both targets, a first order instrumental polarization correction was computed within an annulus extending from $r=10$ to $r=30$ pixels ($\sim 0\farcs{}16 - 0\farcs{}49$). While significant polarized disk signal is present here, a) this is the case throughout the \ac{charis} \ac{fov} for these targets, and b) this approximation is reasonable only with full azimuthal coverage over the annulus, which occurs for $r \lesssim 30$ pixels ($\sim 0\farcs{}5$) with \ac{charis} \ac{pdi} data. However, we note that the inclusion of disk material in the annulus means that this correction is very likely over-estimating the contribution of instrumental polarization; this correction is used only as a placeholder until the formal Mueller Matrix correction is available.
       
    \subsection{AB Aurigae}
\begin{figure}\includegraphics[width=\textwidth]{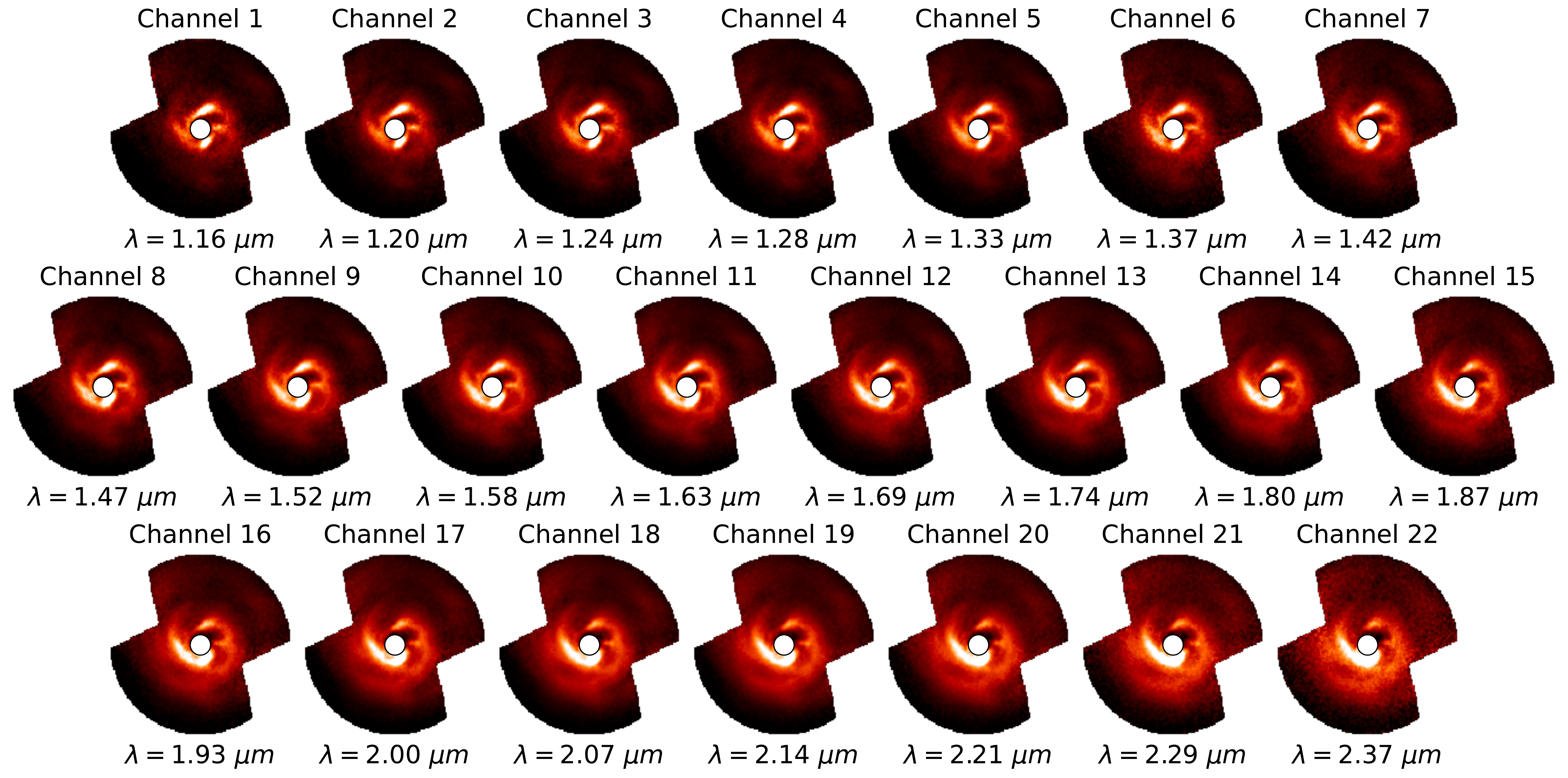}
        \caption{Polarized intensity in 22 wavelength channels for the final, sequence-combined \ac{charis} \ac{pdi}-mode observations of AB Aur. The color stretch is linear throughout, with range chosen arbitrarily for each channel in order to maximize the visibility of disk structures. North is up and east is left, with the \ac{fov} extending to $r\sim 1\farcs{}15$.
        \label{fig:pi_channels}
        }
    \end{figure}

\begin{figure}\includegraphics[width=\textwidth]{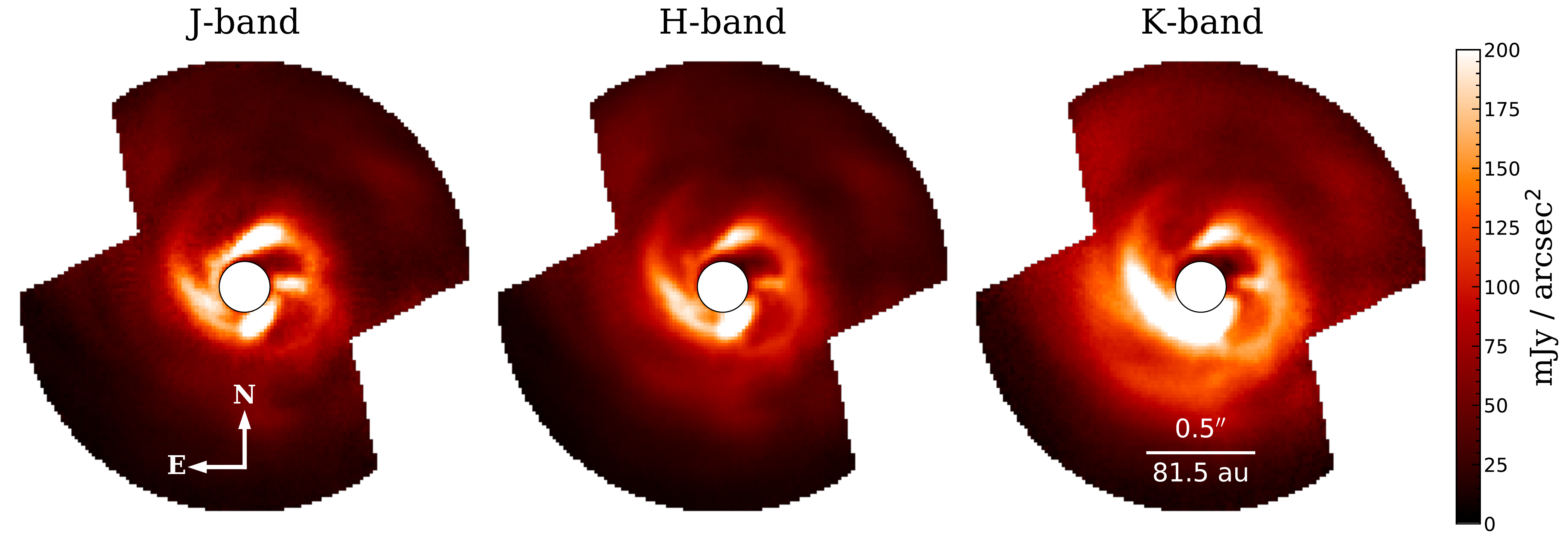}
        \caption{AB Aur \ac{charis} \ac{pi} images created by binning sequence-combined $Q$ and $U$ image cubes along the wavelength axis to approximate \ac{nir} $J$-band (channels $1-5$, $1.16-1.33$ $\mu m$), $H$-band (channels $8-14$, $1.47-1.80$ $\mu m$) and $K$-band (channels $16-21$, $1.93-2.29$ $\mu m$) images. These $Q$ and $U$ images are then combined to produce \ac{pi} using equation \ref{eq:pi}. \label{fig:pi_jhk}
        }
    \end{figure}

\begin{figure}
    {\includegraphics[width=0.495\textwidth]{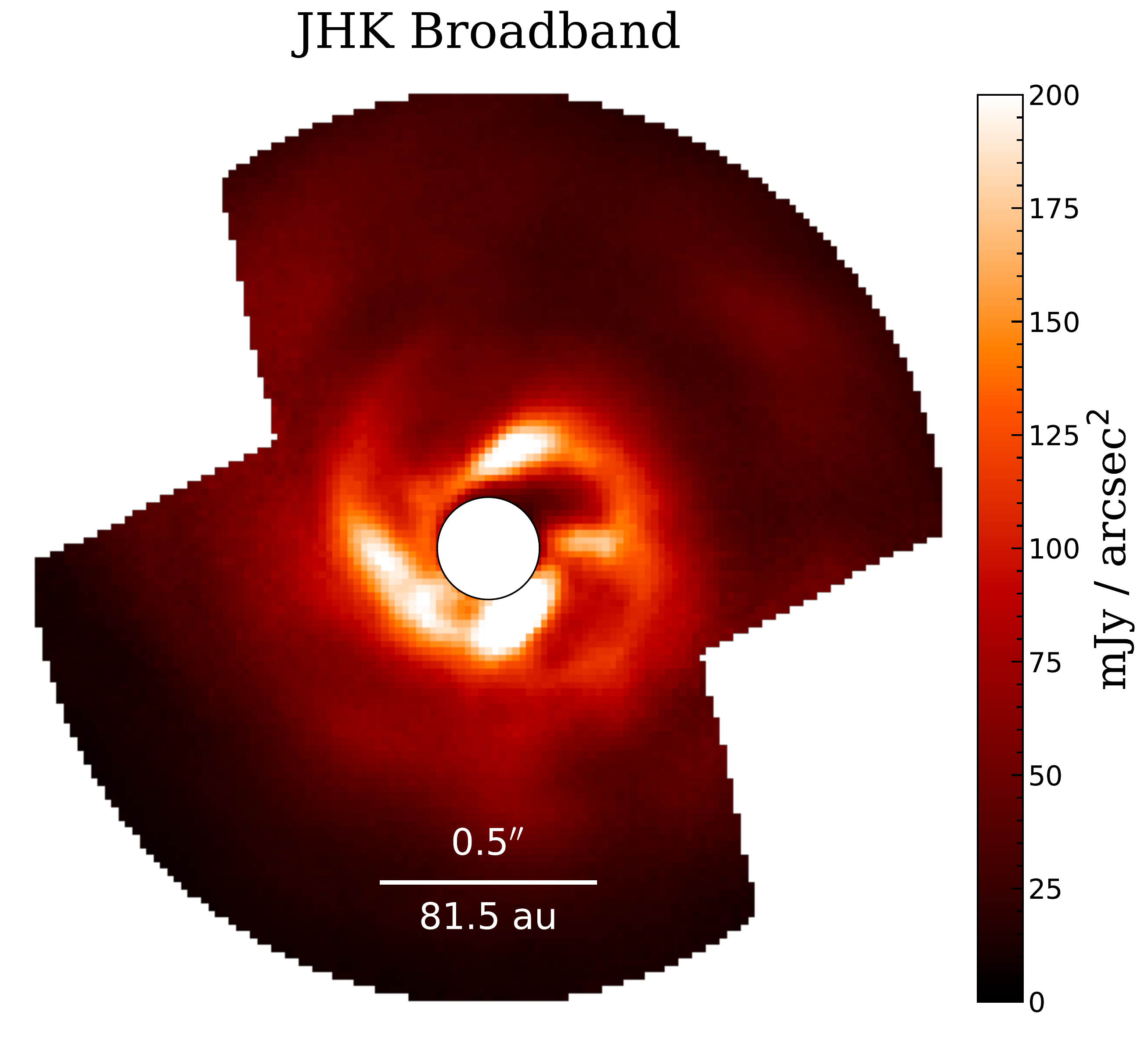}}
    \hfill
    {\includegraphics[width=0.495\textwidth]{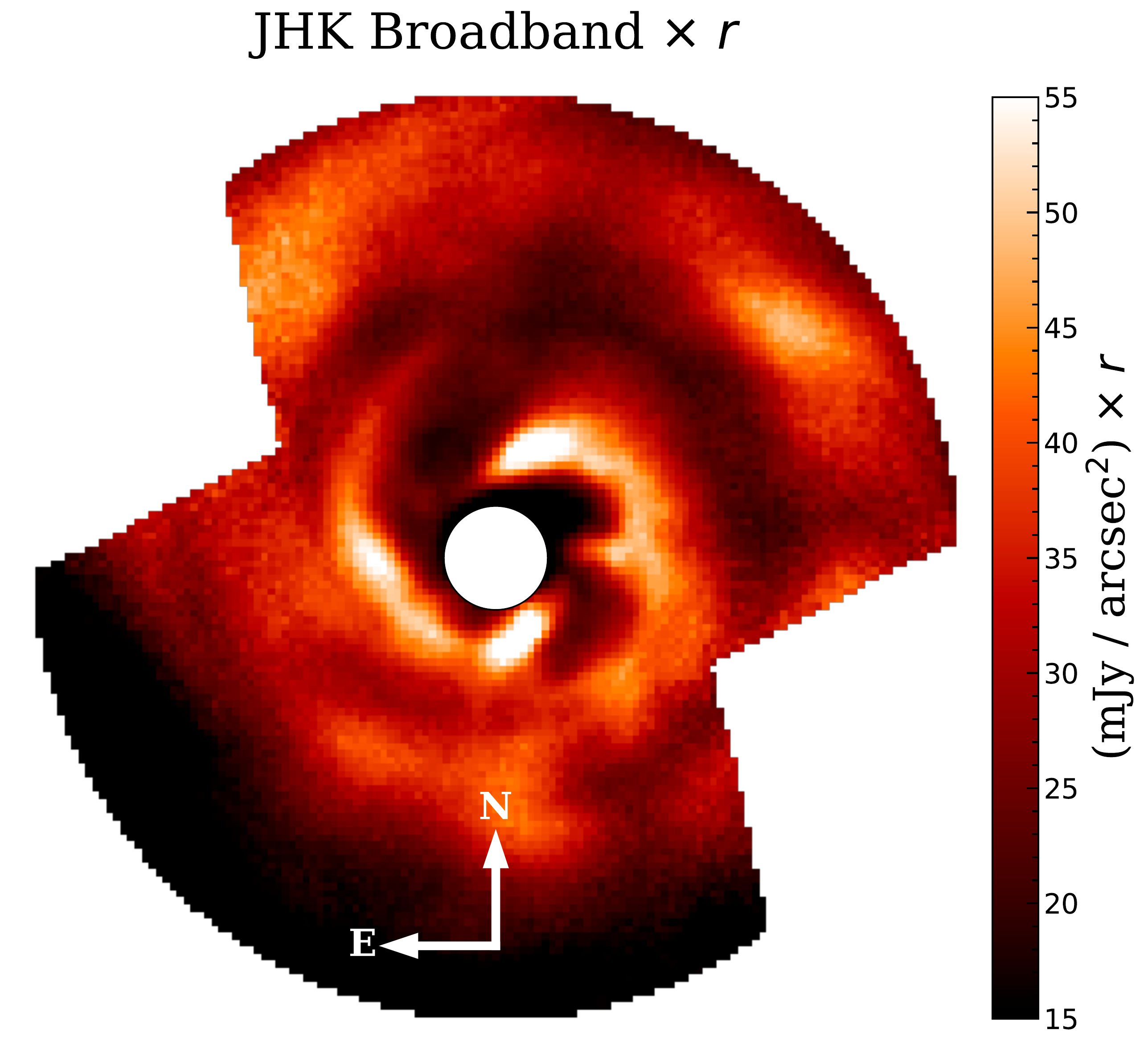}}
    \caption{\ac{charis} wavelength-collapsed \ac{pi} image of AB Aur created by combining all wavelength channels of the $Q$ and $U$ image cubes before application of equation \ref{eq:pi}. The image on the left is presented as the images of Figure \ref{fig:pi_jhk}, while the image on the right has been multiplied by the projected stellocentric separation in arcsec (assuming an inclination of $30\degr{}$ and a position angle of $60\degr{}$). \label{fig:pi_bb}
    }
    \end{figure}

    Following preprocessing, AB Aurigae observations were matched based on \ac{hwp} position to produce 16 complete cycles. We then applied the double-differencing \ac{pdi} technique of Section \ref{sec:pdi} to produce: a) a sequence-combined \ac{pi} image cube (Figure \ref{fig:pi_channels}), b) J, H, and K-band \ac{pi} images (Figure \ref{fig:pi_jhk}), and c) a wavelength collapsed broadband \ac{pi} image (Figure \ref{fig:pi_bb}). These results show unambiguous detections of the complex spiral-armed disk of AB Aur -- not only in wavelength-collapsed products, but also in all 22 individual wavelength channels.
    
    \subsection{TW Hydrae} 
\begin{figure}\includegraphics[width=\textwidth]{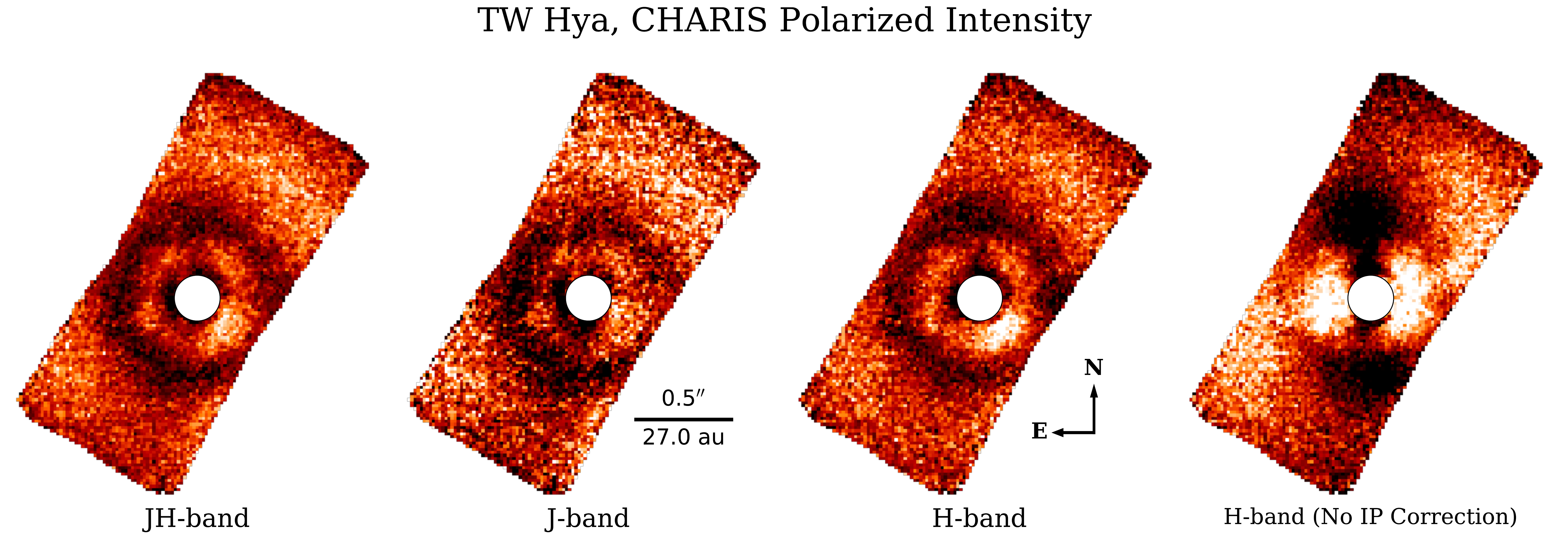}
        \caption{TW Hya \ac{charis} \ac{pi} broadband images created by binning sequence-combined $Q$ and $U$ image cubes along the wavelength axis before application of equation \ref{eq:pi}. The leftmost image, labeled $JH$-band, is created by combining both $J$-band (channels $1-5$, $1.16-1.33$ $\mu m$) and $H$-band channels ($8-14$, $1.47-1.80$ $\mu m$). The third and fourth images provide a comparison of the result in $H$-band with and without first order instrumental polarization correction. The correction is necessary for a clear detection in this case.
        \label{fig:twhya_pi}
        }
    \end{figure}

    \ac{hwp} matching for TW Hya results in 11 complete cycles. Following the double-differencing procedure, we mask any pixels with coverage in less than $25\%$ of derotated images to mitigate noisy edges in the products. Figure \ref{fig:twhya_pi} shows \ac{pi} images for TW Hya, including a comparison of the $H$-band \ac{pi} products with and without first-order instrumental polarization correction. $K$-band is not included; the lack of sky frames for this night combined with the weaker \ac{ao} correction resulting from the target's faintness produce only a marginal detection of disk features in $K$-band.
    
\section{CHARIS PDI Disk Forward Modeling}\label{sec:fwdmodeling}
    For other varieties of differential imaging (e.g. \ac{adi}, \ac{rdi}, \ac{sdi}), detailed and computationally intensive forward-modeling is needed to quantify the (typically significant) erroneous attenuation of disk signal that occurs during nulling of the stellar \ac{psf}. For \ac{pdi}, the observed disk image is generally much closer to the true disk, and forward-modeling is relatively simple. 
    
    For this purpose, we consider two factors affecting the resulting disk signal:
    \begin{enumerate}
        \item the \ac{psf} during our observations
        \item the slight difference in orientation between $X^+$ and $X^-$ used in double differencing (see Equation \ref{eq:double_diff})
    \end{enumerate}
    
    To simulate these effects for a disk model, we proceed as follows. First, synthetic $Q$ and $U$ images are generated for the model. These images are then rotated to match the array of parallactic angles of the appropriate $X^\pm$ data (i.e. those of the utilized images at \ac{hwp} angles of 0$\degr{}$ and $45\degr{}$ for $Q$ or $22\fdg{}5$ and $67\fdg{}5$ for $U$). This results in two three-dimensional sets of model image sequences. Each frame is then convolved with the empirical \ac{psf} model for each wavelength channel (created from either the satellite spots or the unocculted stellar \ac{psf} in the data), and frames corresponding to $X^-$ are multiplied by $-1$. Finally, values falling beyond the \ac{charis} \ac{pdi}-mode \ac{fov} are set to zero. This produces a sequence of image cubes for $Q^\pm$ and $U^\pm$ comparable to the polarized astrophysical signal contained in the data following single-differencing (Equation \ref{eq:single_diff}). 
    
    Once synthetic data sequences containing only disk signal have been created, the double-differencing procedure is applied (Equation \ref{eq:double_diff}), using the same \ac{hwp} cycle matches as for the data. Since the orientations of the model images are offset by the same amount as the real data, any effects that result from this will be propagated to the final product from forward-modeling.  Following this, the resulting sequences are derotated and combined in the same manner as the data.
    
\begin{figure}\includegraphics[width=\textwidth]{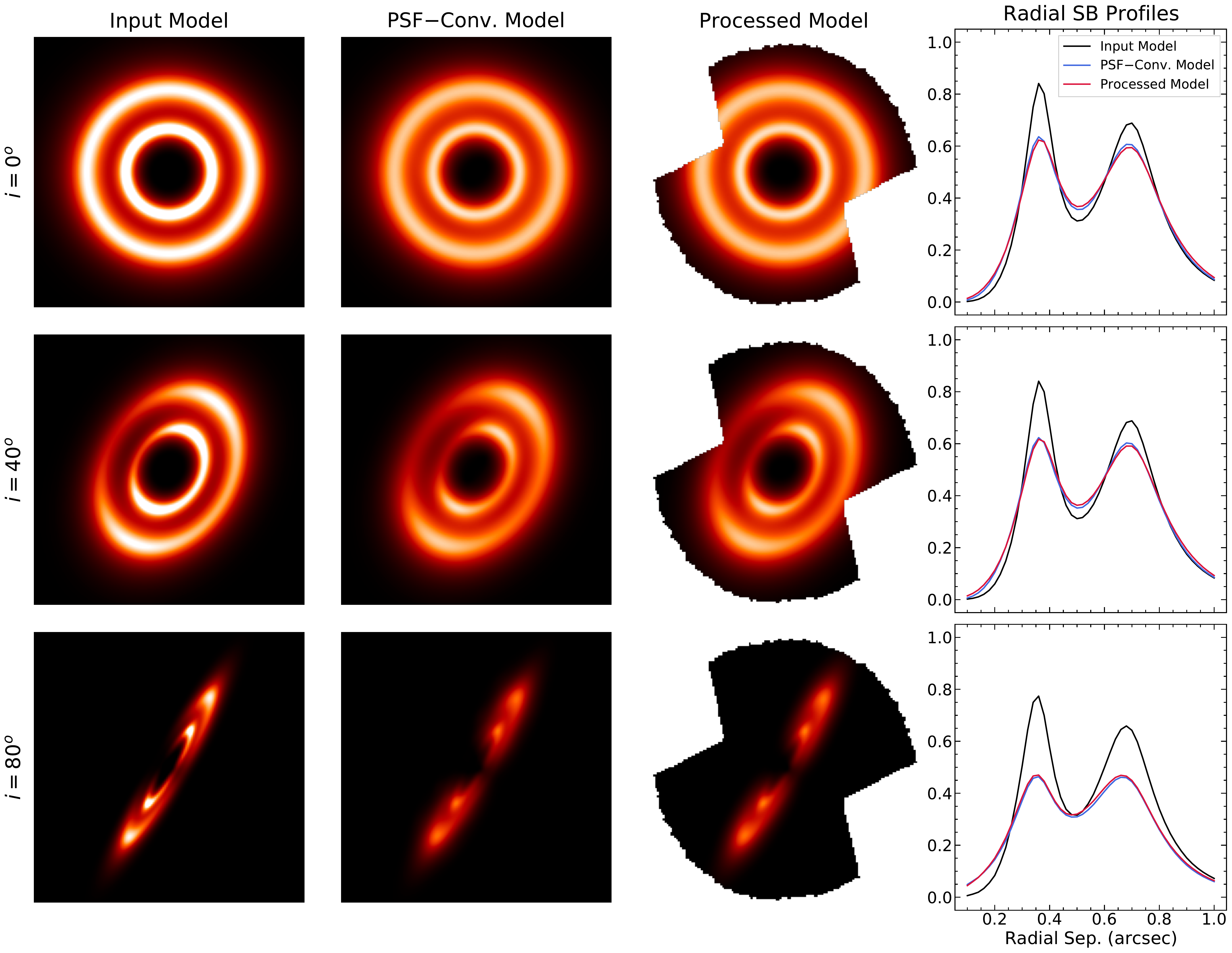}
        \caption{A set of simple synthetic scattered light disk models in polarized intensity are shown throughout the forward modeling procedure for \ac{charis} \ac{pdi}-mode. For this purpose, we use the \ac{psf}, parallactic angles, and \ac{hwp} matches from the observations of AB Aur. Each row corresponds to a distinct input model of differing inclination. The final column displays the radial \ac{sb} of the images at each stage, measured in apertures of 1 \ac{fwhm} in diameter and along the major axis of the model. \label{fig:fwd_modeling}
        }
    \end{figure}

    Examples of input models alongside corresponding \ac{psf}-convolved models and fully processed models resulting from this procedure are shown in Figure \ref{fig:fwd_modeling}. These results show little difference between the \ac{psf}-convolved and fully processed models. This suggests that, even with the relatively high rate of field rotation for the utilized data, combining exposures at slightly different parallactic angles in double differencing has little effect on the products. This is true even in the case of fine azimuthal features -- such as the highly inclined disk model --  where we should expect the impact to be most significant. Additionally, while we expect no change as a result of the parallactic angle difference in the case of the face-on disk, the change between the \ac{psf}-convolved and processed \ac{sb} profiles for this model appears comparable with the change for inclined models. Based on this, it seems that the changes we see are predominantly from other sources (e.g. interpolation during image rotation). In general, the impact of the offset in parallactic angle between double-difference pairs will be negligible, but may warrant consideration when observing targets very near zenith.
    
    The relative simplicity of disk forward modeling for \ac{pdi} data confers a significant strength for disk studies; since the procedure is much less computationally expensive -- requiring only image rotation, \ac{psf} convolution, and simple algebra --  disk models can be forward-modeled orders of magnitude more quickly than for other disk imaging techniques. Combined with swift optimization algorithms, such as differential evolution \acused{de} (DE; see e.g. Ref \citenum{Lawson2020}), \ac{pdi} data enables a much more detailed analysis of disk morphology and composition.  This is especially so in the case of debris disks, which can be reasonably approximated with simple scattered-light models (rather than more expensive radiative transfer models). Along with the unique ability of \ac{charis} to conduct high contrast polarimetry simultaneously at an array of wavelengths, the inherent strength of \ac{pdi} for disk studies is further magnified. 

\section{Summary and Conclusions}\label{sec:conclusion}
    The recently-introduced spectropolarimetry mode for the Subaru Telescope's \ac{scexao}/\ac{charis} provides an exciting new tool for groups studying planet-forming circumstellar disks. This new observing mode enables simultaneous polarized intensity imaging at an array of \ac{nir} wavelengths in a $1\arcsec \times 2\arcsec$ \ac{fov}. We have summarized available instrument configurations and outlined the general procedure used to process \ac{charis} \ac{pdi} data. In this procedure, we leverage proven existing software for classical-mode \ac{charis} data processing alongside purpose-built tools for carrying out \ac{hwp} matching and double-differencing procedures. With the upcoming implementation of a full instrumental polarization correction \cite{vanHolstein2020}, this new observing mode for \ac{scexao}/\ac{charis} will enable numerous exciting avenues of study relevant to planet-forming disks.
    
    Forward-modeling for disks in \ac{pdi} data, using the simple procedure outlined in Section \ref{sec:fwdmodeling}, is extremely time efficient compared to the procedures for other differential imaging techniques. As such, \ac{charis} \ac{pdi}-mode will be uniquely well-suited to studying parameters accessible through forward modeling that are expected to have wavelength dependence, such as scattering phase function.
    
    \ac{charis} \ac{pdi}-mode will also have utility in helping to understand the sources of spiral arms in disks which exhibit them. From continued observation of such disks, the movement of the features can be measured, which can then be used to discriminate between their two likely causes: yet-unseen companions and gravitational instability \cite{Ren2020}. While total intensity observations might also enable such investigations, it would be much more challenging given that the amount and location of disk attenuation will likely vary between epochs. Being free of significant attenuation, \ac{pdi} data will significantly strengthen such investigations.
     
    Finally, comparison of polarized intensity and total intensity \ac{pdi} products (e.g. from application of \ac{rdi}) enables measurement of fractional polarization ($\textit{PI}/I$) across \ac{charis}'s array of wavelengths. These measurements can then be used to identify significant sources of thermal emission relative to other parts of the disk, such as embedded protoplanets, as well as to inform us regarding the properties of the dust within the disk, such as porosity \cite{Hughes2018}.

\acknowledgments    
 
This research is based on data collected at Subaru Telescope,
which is operated by the National Astronomical Observatory of Japan. We are honored and grateful for the opportunity of observing the Universe from Maunakea, which has cultural, historical and natural significance in Hawaii.

We wish to acknowledge the critical importance of the current and recent Subaru telescope operators, daycrew, computer support, and office staff employees. Their expertise, ingenuity, and dedication is indispensable to the continued successful operation of Subaru.  

The development of SCExAO was supported by the Japan Society for the Promotion of Science (Grant-in-Aid for Research \#23340051, \#26220704, \#23103002, \#19H00703 \& \#19H00695), the Astrobiology Center of the National Institutes of Natural Sciences, Japan, the Mt Cuba Foundation and the director’s contingency fund at Subaru Telescope.  We acknowledge funding support from the NASA XRP program via grants 80NSSC20K0252 and NNX17AF88G.  T.C. was supported by a NASA Senior Postdoctoral Fellowship.

This work has made use of data from the European Space Agency (ESA) mission {\it Gaia} (\url{https://www.cosmos.esa.int/gaia}), processed by the {\it Gaia} Data Processing and Analysis Consortium (DPAC,
\url{https://www.cosmos.esa.int/web/gaia/dpac/consortium}). Funding for the DPAC has been provided by national institutions, in particular the institutions participating in the {\it Gaia} Multilateral Agreement.

\bibliography{report} 
\bibliographystyle{spiebib} 

\end{document}